\documentclass[10pt,english, aps,superscriptaddress,showpacs,floatfix,prl,lengthcheck,floatfix,nofootinbib,twocolumn]{revtex4-1}

\usepackage[T1]{fontenc}
\usepackage[latin9]{inputenc}
\usepackage{amsmath}
\usepackage{graphicx}

\makeatletter

\usepackage[latin9]{inputenc} 

\usepackage{braket}

\usepackage[T1]{fontenc}
\usepackage{color}
\usepackage{babel}
\usepackage{verbatim}
\usepackage{amssymb,amsfonts,amsmath}
\usepackage{graphicx}
\usepackage[pdfusetitle,colorlinks,linkcolor=blue,citecolor=blue,urlcolor=blue]{hyperref}
\usepackage{microtype}

\makeatother

\usepackage{babel}
\begin{document}
\title{Formation, stability, and highly nonlinear optical response of excitons
to intense light fields interacting with two-dimensional materials
\\
}
\author{Eduardo B. Molinero}
\email{ebmolinero@gmail.com}

\email{e.b.molinero@csic.es}

\affiliation{\emph{Instituto de Ciencia de Materiales de Madrid (ICMM), Consejo
Superior de Investigaciones Científicas (CSIC), Sor Juana Inés de
la Cruz 3, 28049 Madrid, Spain }}
\author{Bruno Amorim}
\affiliation{\emph{Centro de Física das Universidades do Minho e do Porto (CF-UM-UP)
and Laboratory of Physics for Materials and Emergent Technologies
(LaPMET), Universidade do Minho, 4710-057 Braga, Portugal}}
\author{Mikhail Malakhov}
\affiliation{\emph{Departamento de Química, Universidad Autónoma de Madrid, 28049
Madrid, Spain}}
\author{Giovanni Cistaro}
\affiliation{\emph{Departamento de Química, Universidad Autónoma de Madrid, 28049
Madrid, Spain}}
\author{Álvaro Jiménez-Galán}
\affiliation{\emph{Instituto de Ciencia de Materiales de Madrid (ICMM), Consejo
Superior de Investigaciones Científicas (CSIC), Sor Juana Inés de
la Cruz 3, 28049 Madrid, Spain }}
\author{Misha Ivanov}
\affiliation{\emph{Max Born Institute, Max-Born-Straße 2A, 12489, Berlin, Germany}}
\affiliation{\emph{Department of Physics, Humboldt University, Newtonstraße 15,
12489 Berlin, Germany}}
\affiliation{\emph{Blackett Laboratory, Imperial College London, London SW7 2AZ,
United Kingdom}}
\affiliation{\emph{Technion -- Israel Institute of Technology, 3200003 Haifa,
Israel}}
\author{Antonio Picón}
\affiliation{\emph{Departamento de Química, Universidad Autónoma de Madrid, 28049
Madrid, Spain}}
\author{Pablo San-José}
\affiliation{\emph{Instituto de Ciencia de Materiales de Madrid (ICMM), Consejo
Superior de Investigaciones Científicas (CSIC), Sor Juana Inés de
la Cruz 3, 28049 Madrid, Spain }}
\author{Rui E. F. Silva}
\email{ruiefdasilva@gmail.com}

\email{rui.silva@csic.com}

\affiliation{\emph{Instituto de Ciencia de Materiales de Madrid (ICMM), Consejo
Superior de Investigaciones Científicas (CSIC), Sor Juana Inés de
la Cruz 3, 28049 Madrid, Spain }}
\affiliation{\emph{Max Born Institute, Max-Born-Straße 2A, 12489, Berlin, Germany}}
\begin{abstract}
Excitons play a key role in the linear optical response of 2D materials.
However, their significance in the highly nonlinear optical response
to intense mid-infrared light has often been overlooked. Using hBN
as a prototypical example, we theoretically demonstrate that excitons
play a major role in this process. Specifically, we illustrate their
formation and stability in intense low-frequency fields, where field
strengths surpass the Coulomb field binding the electron-hole pair
in the exciton. Additionally, we establish a parallelism between these
results and the already-known physics of Rydberg states using an atomic
model. Finally, we propose an experimental setup to test the effect
of excitons in the nonlinear optical response
\end{abstract}
\maketitle

\section*{Introduction}

Emergence is a fundamental concept in condensed matter physics \citep{Anderson1972}.
Pertinent examples include superconductivity \citep{Keimer2015,Fernandes2022},
magnetism \citep{Huang2020}, and topological phases \citep{hasan2010,qi2011}.
Another example are excitons \citep{Koch2006}: the quasiparticles
created by attraction between an electron excited to the conduction
band and the hole left in the valence band. In 2D materials excitons
have particularly significant binding energy $\ensuremath{E_{{\rm bind}}\sim1}$
eV and play dominant role in their linear optical response \citep{wang2018colloquium}.
What should one expect for the highly nonlinear optical response,
when 2D solids interact with intense low-frequency laser fields? 

This is a highly pertinent key question for high harmonic generation
(HHG) in solids, which has emerged as an important direction in ultrafast
condensed matter physics \citep{kruchinin2018,Ghimire2018,Goulielmakis2022}.
Will excitons provide strong contribution to intense-field driven
nonlinear response, or will they simply not survive the strong laser
field, which typically exceeds the Coulomb field that binds the electron
and the hole together? Answering this question is important both fundamentally
and for applications. At the fundamental level, HHG offers a unique
window into the electronic structure and dynamics in trivial, topological,
and strongly correlated solids far from equilibrium \citep{Ghimire2010,Vampa2015,Luu2015,hawkins2015,wismer2016_yako,tancgone-dejean2017_bandstructure,tancogne-dejean2018_uhubbard,murakami18,Silva2018,Silva2019,chacon20,Uzan2020,Orthodoxou2021,Schmid2021,Ortmann2021,Alcala2022_hightc,Bharti2022,alshafey2022,molinero23}.
Interpreting these dynamics without understanding the role of excitons
is hardly adequate. For applications, if excitons can generate bright
high harmonics, their role would be important when using high harmonics
as bright solid-state sources of ultrashort VUV-XUV radiation\citep{Vampa2017,nourbakshs2021_merdji}. 

While the entry of intense light fields into condensed matter physics
is relatively recent \citep{Ghimire2010}, interaction with such fields
in general and HHG in particular have been extensively studied in
atoms \citep{krausz2009}. Rydberg states, the atomic analogues of
excitons \citep{haug2009quantum}, were found to play a surprisingly
important role in strong-field ionization from the ground state (the
atomic analogue of electron injection into the conduction band.) Prominent
examples are the so-called frustrated tunnelling \citep{Yudin2001,nubbemeyer08,eichmann2013,Eichmann2023}
and the Freeman resonances in multi-photon ionization \citep{freeman1987,Eichmann2023}.
The remarkable stability of Rydberg states against intense laser fields,
predicted in \citep{Fedorov1989,Fedorov1990,Fedorov2012}, was confirmed
in \citep{deBoer1993_hgmuller,deBoer1994_hgmuller}, dramatically
demonstrated in \citep{eichmann2013}, and even led to lasing during
laser filamentation in dense nitrogen gas \citep{Richter2020}. Multiphoton
Rydberg excitations have been found to contribute to harmonic emission
during the laser pulse \citep{Toma1999,beaulieu2016,mayer2022} and
free induction decay after its end \citep{beaulieu2017_mairesse,bengs2022}.
In contrast, apart from the pioneering works \citep{tancogne-dejean2018_uhubbard,ikemachi2018,udono2022,jensen2023},
the role of excitons in the strong field regime has been generally
ignored, with in-depth analysis of their dynamics and the physics
of their creation and destruction in strong laser fields lacking until
now. 

In this work we aim to fill this gap. We show that high harmonic emission
can reveal the formation of excitons by both significantly increasing
both the overall harmonic yield, by about an order of magnitude, and
the emission intensity at energies near excitonic excitations, by
an even stronger two orders of magnitude. We also show that shifting
the exciton binding energy by using a substrate provides a tell-tale
sign of their contributions. Time-resolved analysis of the emission
shows the formation dynamics and the remarkable stability of excitons
against strong light fields. In spite of the emergent, interaction-induced
nature of excitons, we consistently find strong similarity in their
strong field dynamics with that of single-particle Rydberg states.
This connection highlights how a non-trivial emergent quasiparticle,
such as an exciton in a sea of interacting electrons, can behave much
like a single-particle excitation in an atomic gas, even when driven
by an intense field.

\section*{Results}

\begin{figure}
\begin{centering}
\includegraphics[width=1\columnwidth]{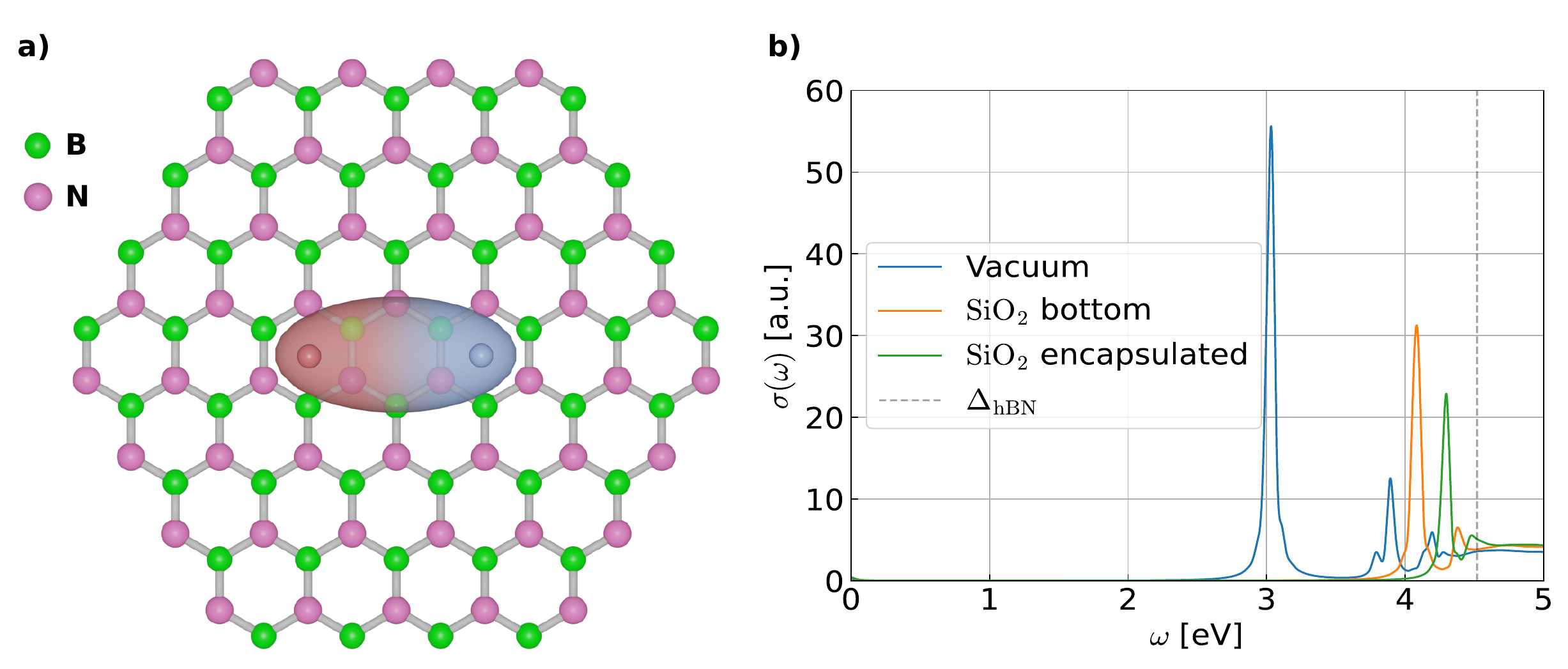}
\par\end{centering}
\caption{\label{fig:linear_diagram}(a) Crystalline structure of hBN alongside
of a depiction of an exciton. (b) Absortion spectra of hBN in terms
of different substrates .}
\end{figure}

The nonlinear optical response of excitons in 2D materials can be
simulated using real-time equation of motion techniques. Here we perform
simulations using hexagonal boron nitride (hBN), see Methods section.
The choice of this specific material was based on two reasons: it
is a prototypical 2d semiconductor \citep{Roldn2017} and it is known
to host excitonic states with large binding energies \citep{galvani2016,Ferreira2019}.
Furthermore, one of the key features is the possibility of engineering
the effect of interactions by changing the substrate. Introducing
a substrate with a higher dielectric constant effectively screens
electronic interactions, thereby reducing the energy required to bind
the electron-hole pair \citep{Hsu2019,NguyenTruong2022} ,i.e., by
shifting the first excitonic state closer to the conduction band.
This behaviour can be appreciated by looking at the absorption spectrum
shown in Fig.\ref{fig:linear_diagram}b (for further information see
Methods).

To see such high harmonic response, we considered a laser pulse in
the mid-infrared regime ($3\,\mu$m) with an intensity of $1.16\,$
TW/cm$^{2}$, a total duration of $20$ optical cycles and with a
$\cos^{2}$ envelope. The field will be oriented in the $\Gamma-K$
direction. However, we must note that the effects here described are
robust against variations of the parameters.

\begin{figure*}
\begin{centering}
\includegraphics[width=2\columnwidth]{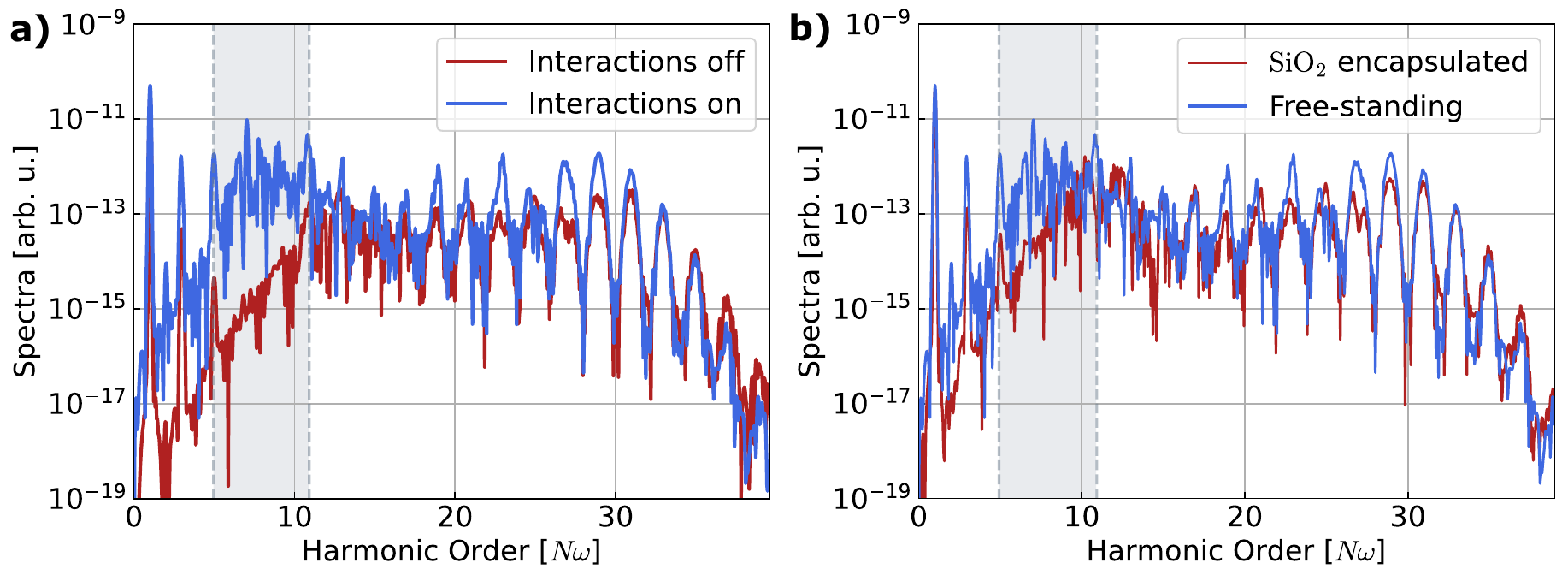}
\par\end{centering}
\caption{\label{fig:hbn}(a) High harmonic generation spectrum computed for
a monolayer of hBN with (blue colour) and without (red colour) electronic
interactions. (b) HHG spectrum computed for a monolayer of free-standing
hBN (blue colour) and encapsulated in $\text{SiO}_{2}$ (red colour).
The spectrum is obtained for a laser pulse in the $\Gamma-K$ direction
along its parallel direction.}
\end{figure*}

As a first step, we have performed two calculations, one where excitons
are present and another one where excitonic effects are neglected.
In Figs. \ref{fig:hbn}(a-b), we show the comparison of the high-harmonic
spectrum between the system with and without excitons. In both cases,
they display the general trend of HHG in solids, i.e., the amplitude
of low-order harmonics decays as the order is increased, until the
energy of the harmonics equals the band gap of the material, roughly
at the tenth harmonic. At this point, a plateau of high-harmonics
emerges, but once they reach their cutoff condition (around the 30th
harmonic), an exponential decay of the harmonics ensues \citep{vampa2014theoretical,kruchinin2018}.
Although the qualitative behavior is similar, the most outstanding
difference between the interacting and non-interacting scenarios is
the presence of an enhancement in the intensity between the fifth
and the eleventh harmonic (gray areas in Figs. \ref{fig:hbn}). This
enhancement coincides precisely with the energy of the exciton in
hBN. A less pronounced but also visible difference is the enhancement
of harmonics in the plateau region. 

As mentioned above, excitons are bound states with energies inside
the gap of the single-particle spectrum, situated between the valence
and conduction bands. Thus, the application of a strong laser field
to the system results in a increased probability for electrons to
transition from the valence to the conduction bands, owing to the
availability of additional channels provided by the excitonic states.
In the transition process, excitons thus play the role of stepping
stones across the gap. Therefore, having more channels available to
excite to the valence band will increase the excited electronic population.
This explains the enhancement of the harmonics in the plateau when
including excitons. Moreover, the amplification is primarily concentrated
in the fifth/seventh harmonic, which corresponds to an energy of 2.1/2.9
eV. This energy region is roughly equivalent to difference $\Delta_{{\rm hBN}}-E_{{\rm bind}}$,
between the hBN bandgap $\Delta=4.52$ eV and the binding energy of
the exciton $E_{{\rm bind}}\approx2.0$ eV \citep{galvani2016,ridolfi2020expeditious,cistaro2023_edus}.
In other words, the HHG enhancement produced by the exciton states
happens around the energy required for a valence-band electron to
transition into the first exciton state, which reflects how excitons
open a new transition pathway for HHG in the interacting case.

Although comparing interacting and non-interacting systems may seem
relevant on its own, it is not possible to switch interactions on
and off during experiments. However, electronic interactions can be
screened by introducing a dielectric material on the system, as mentioned
before. By employing a really strong dielectric, it becomes possible
to emulate a system with negligible interactions. In Fig. \ref{fig:hbn}(b)
we show the high-harmonic spectrum for free-standing hBN versus hBN
encapsulated in silica which acts as the strong dielectric. Notice
how the same physics takes place: an enhancement in the intensity
is observed when the interactions are stronger. The reason is the
same; the excitonic states acts as extra channels for the tunneling.
However, the degree of enhancement observed is less pronounced compared
to the cases with/without interactions. This is attributed to the
incomplete screening of interactions by the silica encapsulation,
resulting in the persistence of certain excitonic states (see Fig.
\ref{fig:linear_diagram}b). The tunability of excitons in two-dimensional
materials, facilitated by the substrate, offers an experimental platform
for investigating the effects of such quasiparticles in the high-harmonic
spectrum.

\begin{figure}
\begin{centering}
\includegraphics[width=1\columnwidth]{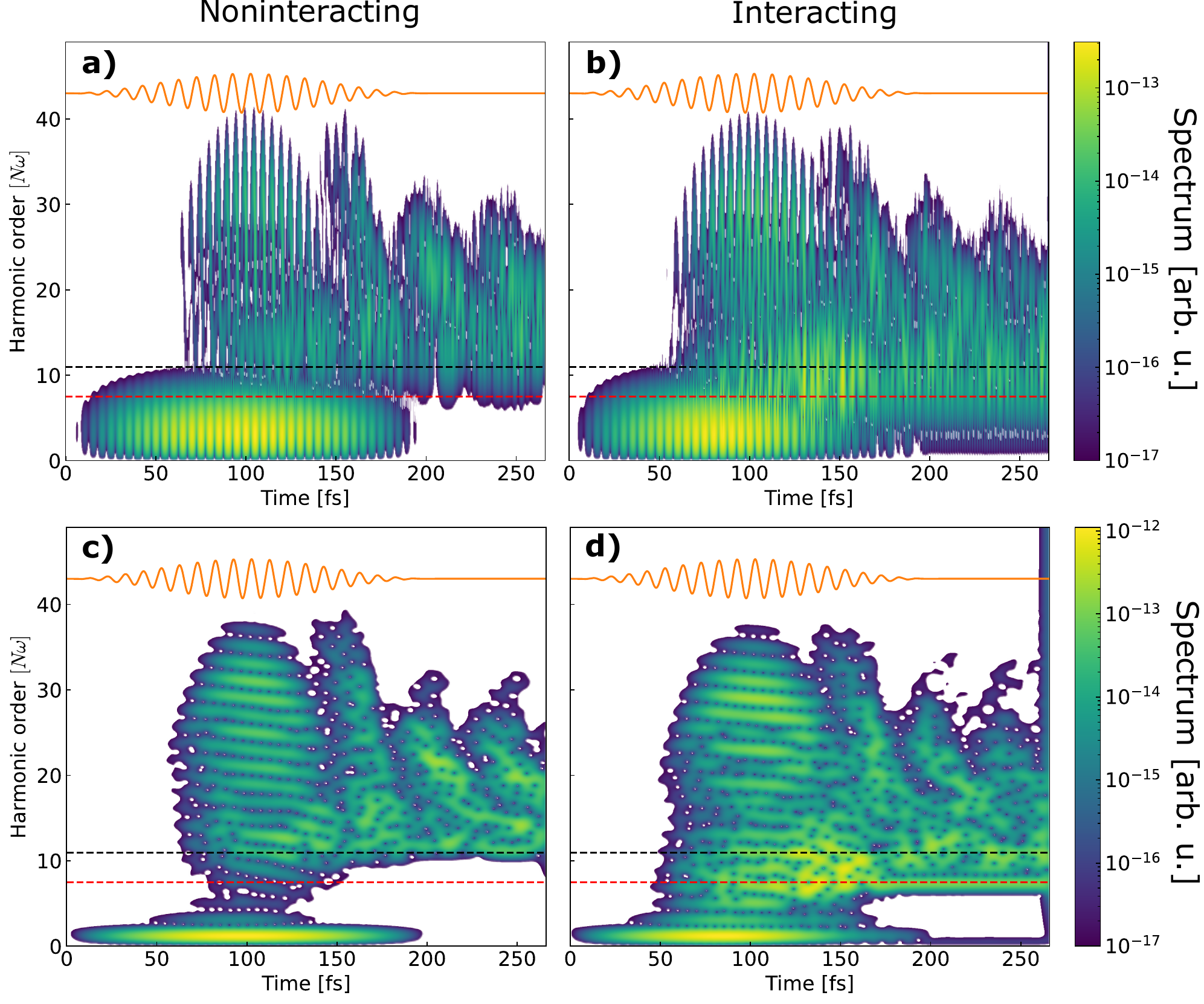}
\par\end{centering}
\caption{\label{fig:gabor}Gabor profile of the harmonic signal for various
cases. The first column corresponds to the non-interacting case while
the second one correspond to the interacting case. Upper row corresponds
to a gaussian window of width $\sigma=(3\omega_{L})^{-1}$ to have
proper resolution in time while the lower row corresponds to a window
of smaller width, $\sigma=(\omega_{L}/2)^{-1}$, to have better resolution
in the frequency domain. The two dashed horizontal lines corresponds
to the energies associated with $\text{\ensuremath{\Delta}}_{\text{hBN}}$(black)
and $E_{{\rm bind}}$ (red) while the orange line depicts the electric
field.}
\end{figure}

We have seen so far, that excitons have a strong influence in the
highly nonlinear optical response. However, as in atoms, should we
expect that excitons are formed and survive after the end of the pulse
with such strong electric fields? To determine whether the exciton
indeed survive to such strong laser pulse, we have computed the Gabor
transform of the current, including times after the pulse has ended.
The first row (Figs. \ref{fig:gabor}a-b) shows the Gabor profile
where one can see a relevant enhancement in the emission below the
band gap (black line) due to the presence of extra channels. An intriguing
new feature observed in the Gabor profile is the appearance of a more
complicated interference pattern when excitons are present in the
system. This phenomenon arises from the influence of bound states
on the semiclassical trajectories. Moreover, clear signatures of exciton
survival after the pulse can be observed. To verify this, we have
performed a Gabor transform with a reduced width, thereby increasing
the resolution in the frequency domain, see Fig. \ref{fig:gabor}c-d.
In these figures, it is evident that in the presence of interactions,
there is free induction decay precisely at the binding energy of the
exciton as the field ramps down ensuring the survival of the exciton
(red line). Remarkably, strong pulses not only create excitons but
also \emph{stabilize} them, just as happens with Rydberg states in
atoms \citep{eichmann2013}. Additionally, following the analogy with
atomic systems, this observation implies that we cannot fully ionize
a two-dimensional system since the electrons remain bound, forming
excitons. Hence, we show evidence of an analog of frustrated ionization
tunneling \citep{Eichmann2023} in solid-state systems.

The presence of bound states between a fixed ground state and continuum
of quasi-free states (the conduction band), bears some resemblance
with the energy spectrum of an atom. Indeed, within a first approximation
excitons are solutions to the Wannier equation \citep{Kittel2004},
which is nothing more than a Schrödinger equation for a centrosymmetric
potential. This observation suggests the possibility that the HHG
spectrum of the interacting semiconductor could be approximately described
using a simple, non-interacting atomic model, where excitons are replaced
by excited states of the atom. In the following we confirm that this
is indeed the case. However, we want to address the opposite question:
can the \emph{whole} system be qualitatively described using an atomic
model?

\begin{figure}
\begin{centering}
\includegraphics[width=1\columnwidth]{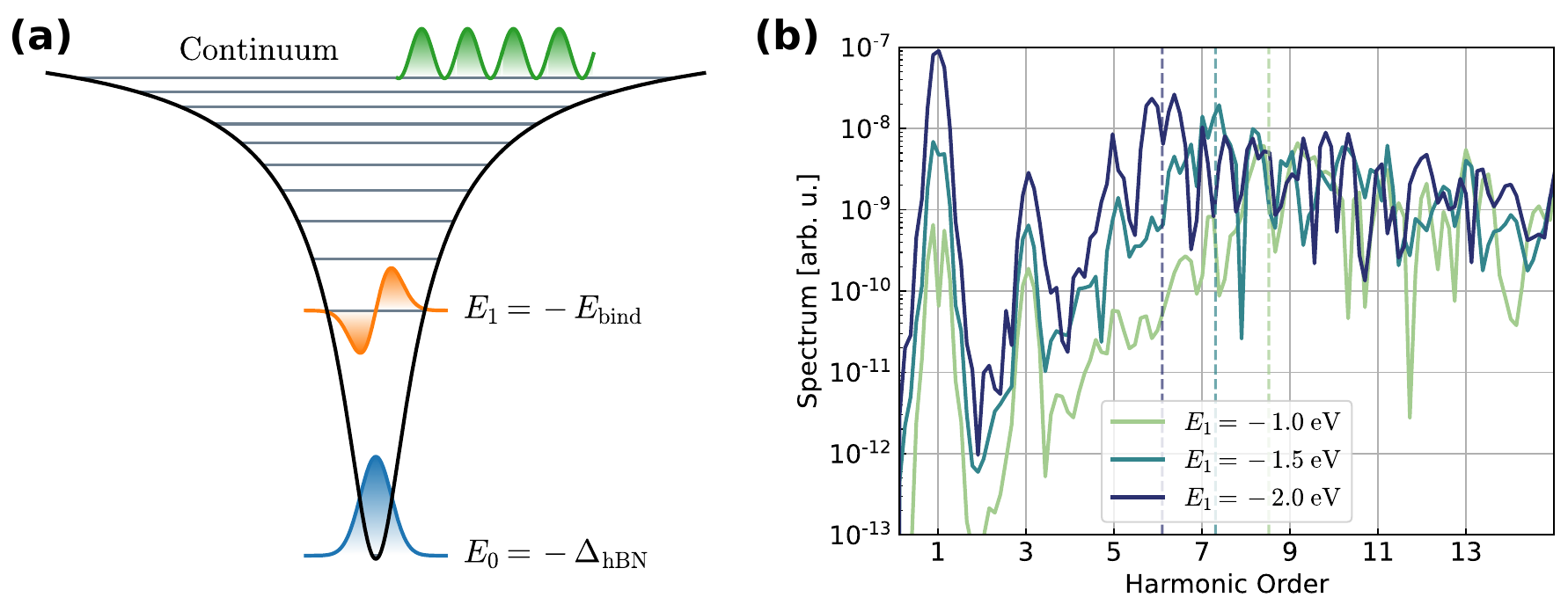}
\par\end{centering}
\caption{\label{fig:atom}(a) Schematic diagram of $V_{\alpha,\beta}(x)$ {[}WIP{]}.
(b) High-harmonic spectrum of the atomic model. Dashed lines denotes
the place were the difference $|E_{0}|-|E_{{\rm 1}}|$ lies for each
$E_{1}$.}
\end{figure}

To answer this question, we have developed a one-dimensional atomic
model (see Methods for more details) that intends to capture the physics
of hBN excitons. The crucial idea is the use of a softcore potential,
\begin{eqnarray}
V_{\alpha,\beta}(x) & = & \frac{\alpha}{\sqrt{x^{2}+\beta^{2}}},\label{eq:softcore}
\end{eqnarray}
which allows us to model the excitation spectrum of hBN. We will tweak
the potential parameters, $\alpha$ and $\beta$, so that the energy
difference between the ground and the first excited state matches
the crucial energy scale $\Delta_{{\rm hBN}}-E_{{\rm bind}}$. More
specifically, we will fix the ground state energy to $E_{0}=-\Delta_{{\rm hBN}}$
and the first excited state to $E_{1}=-E_{{\rm bind}}$ (see the diagram
in Fig. \ref{fig:atom}a). The laser parameters will be the same as
for the hBN case. However, as reaching the tunneling regime in atoms
tends to require higher field intensities \citep{krausz2009}, we
will increase its value up to $4.5$ TW/cm$^{2}$; this particular
value was chosen so that the cutoff in the harmonics were the same
in both systems. Fig. \ref{fig:atom}b shows the HHG spectrum of the
atomic model in terms for various $E_{1}$ energies. The spectrum
displays the typical characteristics of an atomic spectrum \citep{krausz2009}:
the low-order, perturbative harmonics, followed by the plateau harmonics
caused by the recombination processes. It is worth noting that when
the energy of the first excited state $E_{1}$ is raised, the appearance
of the plateau shifts to higher harmonic frequencies. Such shift can
be understood in the same way as for the hBN case: the closer the
first excited is to the ground state, the more likely it is for the
electron to transition into the continuum, thus facilitating the onset
of the plateau in the harmonic spectrum. This is also the kind of
enhancement produced by Rydberg states found in atomic gases \citep{beaulieu2016}.

\begin{figure*}
\begin{centering}
\includegraphics[width=2\columnwidth]{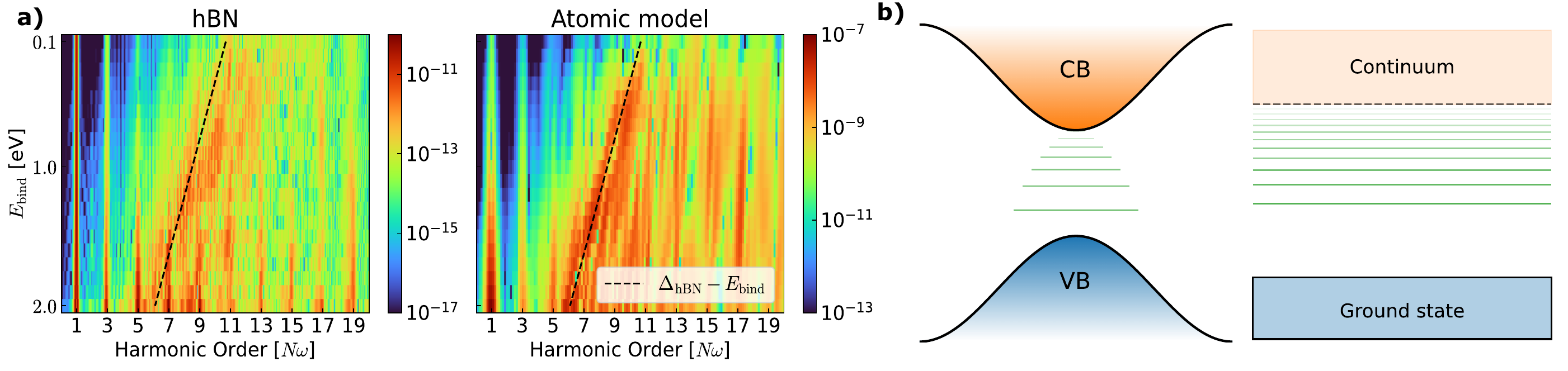}
\par\end{centering}
\caption{\label{fig:comp}(a) High-harmonic spectrum comparison between the
atomic system (left) and the hBN case (right). The dashed black lines
denotes where the line $\Delta_{{\rm hBN}}-E_{{\rm bind}}$ lies.
(b) Schematic diagram to clarify the equivalence between the two systems
{[}WIP{]}.}
\end{figure*}

To better understand the similarities between the two systems, we
conducted a scan encompassing different excitonic energies. Although
the exciton binding energy $E_{{\rm bind}}$ is in principle a fixed
physical quantity (at least if we neglect screening effects from the
electrostatic environment), we can adjust its value from $-2.0$ to
$0.1$ to clarify its effect on the HHG spectrum. This is done by
changing the amplitude of the Rytova-Keldysh potential \citep{cistaro2023_edus}.
For each binding energy we then compute the corresponding parameters
$\alpha,\beta$ of $V_{\alpha,\beta}(x)$. In Fig. \ref{fig:comp}a
we plot the result, comparing the HHG spectrum between the 2d system
and the atomic one in terms of the first exciton binding energies.
Both systems display a qualitatively similar HHG spectrum; the enhancement
is located precisely at the specific harmonic that corresponds to
$\Delta_{{\rm hBN}}-E_{{\rm bind}}$, see the dashed line. The similarity
between Figs. \ref{fig:comp}(a,b) is striking, given that these correspond
to two very different physical problems: one describes the non-linear
electron dynamics of a two-dimensional material with electron-electron
interaction, while the other corresponds to a one-dimensional non-interacting
atom. The common denominator between the two systems is, as mentioned
before, the existence of a fixed ground state separated from a continuum
of states with excitable states between those two (see Fig.\ref{fig:comp}b),
even if their nature (two-particle vs single-particle) is completely
different. There are other obvious differences, such as the existence
of dispersive bands above the gap in the 2D crystal, which can even
be topological. However, the qualitative aspects of electron dynamics
are insensitive to those differences. Fundamentally, the key quantity
that controls the rate of transition \citep{krausz2009,kruchinin2018},
and hence the emission intensity, is the energy of the lowest excitation
$\Delta_{{\rm hBN}}-E_{{\rm bind}}=|E_{0}|-|E_{1}|$ . In this context,
therefore, an interacting two-dimensional crystal can be understood
qualitatively with a non-interacting atomic gas model, where excitons
play the same role as Rydberg states in enhancing the HHG response
\citep{beaulieu2016,mayer2022}.

The strong field approximation (both in atoms and in solid-state systems)
only considers the existence of a ground state coupled to a continuum
of states, without including any excited states \citep{Amini2019}.
Indeed, going further than the strong field approximation leads to
an enhancement in the high-harmonic emission in atoms \citep{Perez-Hernandez09,PrezHernndez09-thesis,zimmerman2017}.
These previous results bolster our interpretation of excitons as analogs
of Rydbergs states in solids. Moreover, it is known that these Rydberg
states can act as a bottleneck to ionization; the electrons gets trapped
into long orbits (Rydberg states) due to the laser field, frustrating
a complete ionization \citep{nubbemeyer08,eichmann2013} of the system.
Analogously, the same phenomena could take place in crystals, namely
that the strong field excitation of electrons to the conduction bands
could be blocked due to the population of excitonic states. 

\section*{Conclusion}

In this work, the effect of many-body interactions on the high-harmonic
spectra of two-dimensional materials has been studied. The results
shows a significant increase in the emission spectra when accounting
for these interactions, which is associated to the presence of excitonic
states within the system. Specifically, the enhancement is located
at the energy difference between the valence band and the first excitonic
state; the excitons act as extra channels for the tunneling processes
from the valence to the conduction band. Furthermore, we proposed
an experimental setup to effectively test the effect of excitons.
Additionally, we have shown that the exciton does survive such strong
pulses, in complete analogy with the physics observed in atoms due
to the Rydberg states. Finally, we developed a one-dimensional atomic
model to gain physical insight. Using this model, we showed the same
qualitative physics appears in both systems: the presence of excited
states between a fixed ground state and a continuum leads to an enhancement
in the high-harmonic spectra. Our work has shown how interactions
affect the high-harmonic generation in two-dimensional materials and
the role of emergent quasiparticles. These results allow us to reinterpret
the excitons as a many-body version of Rydberg states in the strong
field regime, opening the window for the use of the techniques in
atomic strong field physics for imaging and control of these Rydberg
states in the context of condensed matter targets. 

\section*{Methods}

\emph{hBN simulations --} The microscopic response of the system
to the laser field was obtained by numerically solving the Semiconductor
Bloch Equations (SBEs) in the Wannier Gauge \citep{silva2019_wannier,cistaro2023_edus}.
These equations, in atomic units, reads\\
\begin{gather}
\text{i}\partial_{t}\rho_{nm}(\boldsymbol{k},t)=[H^{(0)}(\boldsymbol{k})+\text{\ensuremath{\Sigma}}(\boldsymbol{k},t),\rho(\boldsymbol{k},t)]_{nm}\\
+|e|\boldsymbol{E}(t)\cdot[\boldsymbol{A}(\boldsymbol{k}),\rho(\boldsymbol{k},t)]_{nm}\\
+\text{i}|e|\boldsymbol{E}(t)\cdot\nabla_{\boldsymbol{k}}\rho_{nm}(\boldsymbol{k},t)
\end{gather}
where $H_{nm}^{(0)}(\boldsymbol{k})$ are the non-interacting terms
of the Hamiltonian, $\text{\ensuremath{\Sigma_{nm}}}(\boldsymbol{k},t)$
accounts for the electronic Coulomb interactions, $\boldsymbol{A}_{nm}(\boldsymbol{k})$
are the multiband Berry connection terms and $n,\,m$ refers to the
band indexes. The electronic interaction are incorporated in the dynamics
at the Fock level \citep{ridolfi2020expeditious,cistaro2023_edus}.
More formally, this means that the self-energy is calculated using
\[
\text{\ensuremath{\Sigma_{nm}}}(\boldsymbol{k},t)=-\sum_{\boldsymbol{k'}}V_{nm}(\boldsymbol{k}-\boldsymbol{k'})\left(\rho_{nm}(\boldsymbol{k}',t)-\rho_{nm}^{0}\right).
\]
where the initial state, $\rho_{nm}^{0}=\rho_{nm}(\boldsymbol{k},0)$,
is completely filled for all the states below the Fermi energy. The
subtraction $\rho_{nm}(\boldsymbol{k}',t)-\rho_{nm}^{0}$ is done
to ensure that we not take into account interactions in the equilibrium
state.

The potential, $V_{nm}(\boldsymbol{k}-\boldsymbol{k'})$, reads
\[
V_{nm}(\boldsymbol{k}-\boldsymbol{k'})=\sum_{\boldsymbol{G}}e^{i(\boldsymbol{k}-\boldsymbol{k'}+\boldsymbol{G})\cdot(\boldsymbol{\tau}_{n}-\boldsymbol{\tau}_{m})}V(\boldsymbol{k}-\boldsymbol{k'}+\boldsymbol{G}),
\]
where $\boldsymbol{\tau}_{n}$ are the center of the Wannier orbitals
and $\boldsymbol{G}$ are the vectors of the reciprocal lattice. The
sum over $\boldsymbol{G}$ is done to ensure the periodicity of the
system. Here, $V(\boldsymbol{q})$, is the Fourier transform of the
Rytova-Keldysh potential, which is known to accurately capture screening
and dielectric effects in two-dimensional materials \citep{Keldysh79,cudazzo11}.

For the monolayer hBN, we used the tight-binding model in which only
the $p_{z}$ orbitals are considered \citep{guinea09-review,silva2019_wannier,cistaro2023_edus}.
The hopping parameter, $t_{0}$, was set to $-2.8$ eV and the on-site
energy for the two atomic species was set to $\varepsilon_{B/N}=\pm2.26$
eV. The density matrix was constructed in a $300\times300$ Monkhorst--Pack
grid and it was time-propagated using a fourth-order Runge-Kutta with
a timestep of $dt=0.1$ atomic units (au). Convergence was ensured
for all the numerical parameters. 

\emph{Atomic model --} The atomic model is based on the solution
of the time-dependent Schrödinger equation (TDSE) for a one-dimensional
atom in the presence of a strong laser field. In atomic units, the
TDSE reads 
\begin{equation}
i\partial_{t}\Psi(x,t)=\Big(T_{{\rm kin}}+V_{\alpha,\beta}(x)+E(t)\cdot x\Big)\,\Psi(x,t),\label{eq:tdse-1}
\end{equation}
where $T_{{\rm kin}}$is the electronic kinetic energy, $V_{\alpha,\beta}(x)$
is the softcore potential (Eq. \ref{eq:softcore}) and $E(t)$ is
the electric field. The TDSE was solved numerically using a fourth-order
Runge-Kutta method with a timestep of $dt=0.1$ au. The initial state,
$\Psi(x,0)$, was selected as the ground state of the time-independent
hamiltonian $H=T_{{\rm kin}}+V_{\alpha,\beta}$. The calculations
were performed in a box of length $L=1000$ au with a grid spacing
of $dx=0.25$ au. We checked that convergence was achieved for all
numerical parameters. 

\section*{Data availability}

The data that support the plots within this paper and other findings
of this study are available from the corresponding authors upon reasonable
request.

\section*{Author contributions}

E. B. M., B. A., M.I., P. S. J. and R. E. F. S. developed the idea.
E. B. M. performed the numerical calculations. M. M., G. C. and A.
P. developed the numerical code used for the SBEs. E. B. M. developed
the numerical code for the atomic TDSE. All authors contributed to
analysis of the results. E. B. M., M. I., P. S. J. and R. E. F. S.
wrote the main part of the manuscript that was discussed by all authors. 

\section*{Competing interests}

The authors declare no competing interests.

\section*{Acknowledgments}

E. B. M. and R. E. F. S. acknowledge support from the fellowship LCF/BQ/PR21/11840008
from \textquotedblleft La Caixa\textquotedblright{} Foundation (ID
100010434). This research was supported by Grant PID2021-122769NB-I00
funded by MCIN/AEI/10.13039/501100011033. 

\bibliographystyle{apsrev4-1}
\bibliography{hhg_excitons}

\end{document}